\documentclass[showkeys,preprint]{revtex4-2}
\usepackage{url}
\usepackage[dvipsnames]{xcolor}
\usepackage{amsfonts,amsmath,amssymb,amsthm,mathtools} %
\usepackage{graphicx} %
\usepackage{epstopdf}

\usepackage{float}

\begin{document}

\title{Organization and Dynamics of Chromosomes}

\author{D. Thirumalai}
\email{dave.thirumalai@gmail.com} 
\affiliation{Department of Chemistry, The University of Texas at Austin, Austin, Texas, USA}
\affiliation{Department of Physics, The University of Texas at Austin, Austin, Texas, USA}

\author{Guang Shi}
\affiliation{Department of Chemistry, The University of Texas at Austin, Austin, Texas, USA}

\author{Sucheol Shin}
\affiliation{Department of Chemistry, The University of Texas at Austin, Austin, Texas, USA}

\author{Changbong Hyeon}
\affiliation{Korea Institute for Advanced Study, Seoul, Korea}

\begin{abstract}
How  long threadlike eukaryotic chromosomes fit tidily in the small volume of the nucleus without significant entanglement is just beginning to be understood, thanks to major advances in experimental techniques. Several polymer models, which reproduce contact maps that measure the probabilities that two loci are in spatial contact, have predicted the three-dimensional structures of interphase chromosomes.  Data-driven approaches, using contact maps as input,  predict that mitotic helical chromosomes are characterized by switch in handedness, referred to as ``perversion".  By using experimentally derived effective interactions between chromatin loci in simulations, structures of conventional and inverted nuclei have been accurately predicted. Polymer theory and simulations show that the dynamics of individual loci in chromatin  exhibit subdiffusive behavior but the diffusion exponents  are broadly distributed, which accords well with experiments. Although coarse-grained models are successful, many challenging problems remain, which require the creation of new experimental and computational tools to understand genome biology. 

\end{abstract}

\keywords{contact and distance maps, polymer models, structural heterogeneity, chromosome Statistical Potential, random helix perversion, glass-like dynamics}

\maketitle


\section{Introduction}
\begin{figure}[h]
\centering\includegraphics[width=0.9\textwidth]{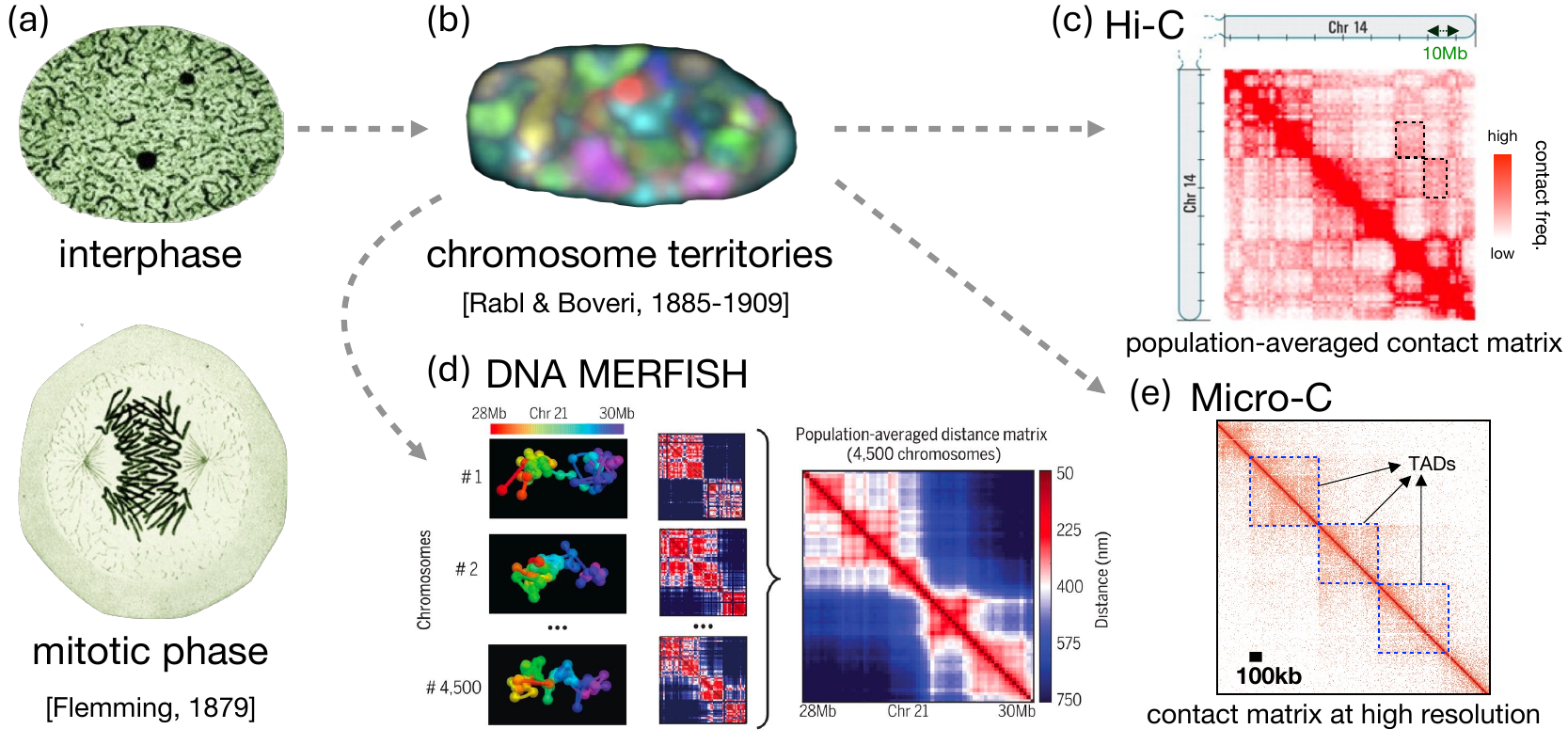}
\caption{ 
(a) Drawings by Flemming, taken from Ref.~\citenum{rieder2003mitosis},  of a cell nucleus in interphase (top) and mitotic phase (bottom).  
(b) Chromosome territories, proposed by Rabl and Boveri, were visualized and confirmed  using DNA fluorescence in situ hybridization (FISH) experiments (image reprinted from Ref.~\citenum{bolzer2005PLoSBiol}).
(c) Experimentally determined contact map for human chromosome 14 ~\cite{lieberman09Science}. Shaded regions along the diagonal correspond to TADs.  The regions enclosed by the black dashed rectangles correspond to compartments.
(d) Multiplexed FISH \cite{wang2016science} combined with super-resolution or diffraction-limited imaging techniques provides 3-dimensional coordinates of chromosomes, which exhibit cell-to-cell structural heterogeneity. The ensemble-averaged distance matrix is in agreement with the Hi-C results. The image is reprinted from Ref.~\cite{bintu2018Science}.
(e) Micro-C \cite{Hsieh2015a,Hsieh2020, Goel2023} reveals chromatin interactions on the length scale of a few kilobase pairs. Regions in blue dashed lines represent TADs.}
\label{fig:fig1}
\end{figure}

It is astonishing that already in 1879, Flemming observed threadlike objects in the mitotic phase that he christened as chromatin \cite{paweletz2001walther}.  He also described subsequent changes in the chromosome structures during different phases of the cell cycle \cite{flemming1882zellsubstanz} (Fig.~\ref{fig:fig1}(a)).  Equally stunning is the introduction of the concept that interphase chromosomes are confined to specific regions in the nucleus by Carl Rabl in 1885 \cite{Rabl1885}. The term, ``chromosome territories" (Fig. \ref{fig:fig1}(b)), referring to this phenomenon, was introduced a decade later by Theodor Boveri \cite{Boveri1909,Cremer10CSHP}. Nearly seventy years would pass, before  the announcement  of the double helix structure of DNA by Watson and Crick \cite{Watson53Nature}, which not only readily explained the replication mechanism but also ushered in the era of molecular biology. About twenty years after the determination of the structure of DNA, Kornberg \cite{Kornberg74Science} showed that the fundamental building block of chromatin is the nucleosome, a protein-DNA complex \cite{Richmond84Nature,Luger97Nature}. These truly landmark discoveries have set the stage for much of the current research in both genome biology and, more recently, the biophysics of genomes \cite{Rowley18NatRevGen,Dekker13NatRevGen,Mirny22CSHL}.  The daunting tasks that remain are  to determine the structures and dynamics  of chromosomes and link them to  many crucial functions that they control, such as gene expression and cell division, which are always carried out by interactions with multiple proteins and motors.   

The discovery that chromatin is made up of repeating units (monomers) of approximately 200 nucleotides wrapped around histone proteins \cite{Kornberg74Science} naturally raised the question of how the  eukaryotic chromosomes consisting of millions of monomers are  packaged in the  small volume of the cell nucleus. (Note that the DNA sequences in the nucleosomes vary greatly.) The quest to answer this major question and to explore the ramifications in biology has gathered momentum in the last two decades thanks to major advances in experimental techniques. The most routinely used method is the chromosome conformation capture \cite{Dekker2002Science} technique that combines chemical cross-linking (first used in 1976 \cite{Hanson76Science}) with deep sequencing \cite{lieberman09Science}  to generate the contact map (Fig. \ref{fig:fig1}(c)), which provides genome-wide data on the  probabilities  that two loci separated by certain genomic distance, are in spatial contact. We will refer to this method and many variants collectively as Hi-C \cite{lieberman09Science,Mirny22CSHL}. Increasingly, direct imaging of sites on chromatin has been used to generate spatial locations of labeled loci on chromosomes \cite{bintu2018Science,Boettiger2016}, with some experiments visualizing up to about 900 loci \cite{Su2020}. Using these methods, contact maps and distance maps (spatial distances between labeled loci) for a number of species have been generated (Fig. \ref{fig:fig1}(d)). These experiments provide  glimpses into the organization of chromosomes, which for practical purposes can be thought of as copolymers consisting of euchromatin or active (A-type) and heterochromatin or inactive (A-type) loci. The contact map on different mammalian cells shows that interphase chromosomes may be described using two major length scales. On the scale of $\sim$ (2--5) Mb (megabase pairs) there are checkerboard patterns \cite{Rao2014Cell} in the contact map, referred to as compartments (Fig. \ref{fig:fig1}(c)). The checkerboard is thought to suggest that stretches of A-type loci lie in the A-compartment while stretches of B-type lie in the B-compartment. Thus, on (2--5) Mb scale A and B phase separate.   On a shorter length scale (a few tens of kilobases to a few Mb), Topologically Associating Domains (TADs) are found as contact-enriched squares along the diagonal in the contact map (Fig. \ref{fig:fig1}(c) and (e)). In other words, loci within a TAD physically interact with greater probability than loci  outside a TAD.

The objective of this perspective is to highlight theoretical and computational methods that have been devised to provide insights into the experimental data to better understand the three-dimensional structures and dynamics of chromosomes.  We show that polymer physics and statistical mechanics concepts are efficacious in uncovering the emergence of the two length scales in chromosome organization. Importantly, we describe methods to predict the unusual structures of chromosomes during cell division. Interestingly, simulations using polymer models show that mammalian chromosomes exhibit glass-like dynamics, which could be understood using theories in the physics of confined polymers.

\section{Computational Methods}
 
Inspired by Hi-C \cite{Dekker2002Science,Dekker13NatRevGen} and imaging experiments  \cite{bintu2018Science,Su2020}, a number of computational models have been introduced to determine the spatial organization of chromosomes. These methods roughly fall under three categories: (1) An energy function relying on polymer representation of chromosomes at different resolutions, with an unknown set of parameters, is assumed to describe the conformations of chromosomes. The unknown parameters are determined by matching the predicted and the experimental contact maps.   (2) In the data-driven methods, experimentally inferred contact maps or imaging data are used to generate an ensemble of three-dimensional (3D) structures without assuming any energy function. The  measured contact maps or mean distances   between the loci are used as constraints to determine the distribution of the 3D coordinates of the chromosomes. (3) A variety of methods based on Machine Learning (ML) have been proposed for genome folding, identification of enhancer--promoter interactions, and other functions \cite{Xie2020, Whalen22NatRevGen,Zhou22NatGen,Fudenberg20NatMethods}. Convolutional neural networks have been used to predict the 3D structures using only the DNA sequence as input \cite{Fudenberg20NatMethods}. The major advantage of sequence-based approaches   is that the effects of mutating single base pairs can be predicted.  ML methods will continue to play an important role in the ability  to understand the nuances of genome organization. Due to space limitations we discuss only polymer and data-driven models here.

\section{Polymer Models}
Polymer model was used nearly thirty years  ago \cite{van92Science,Sachs95PNAS} to explain the fluorescence \textit{in situ} hybridization (FISH) data, which was used to measure the mean-squared distance between as a function of genomic distance. The measurements, performed over a range of 0.15 to 190 Mb, were fit  using a random walk/giant-loop polymer model, which suggested that the loop sizes are on the order of $\approx$ 3 Mb.  However, it is only with the routine generation of contact maps \cite{lieberman09Science,Dekker13NatRevGen,nicodemi2014models}, and imaging data \cite{bintu2018Science}, for a variety of species at various resolutions, that there has been a concerted effort by several groups to develop predictive computational models. We describe three models that cover a spectrum of methods.

\subsection{Strings and Binders Switch (SBS) model}
\begin{figure}[h]
\centering
\includegraphics[width=0.9\textwidth]{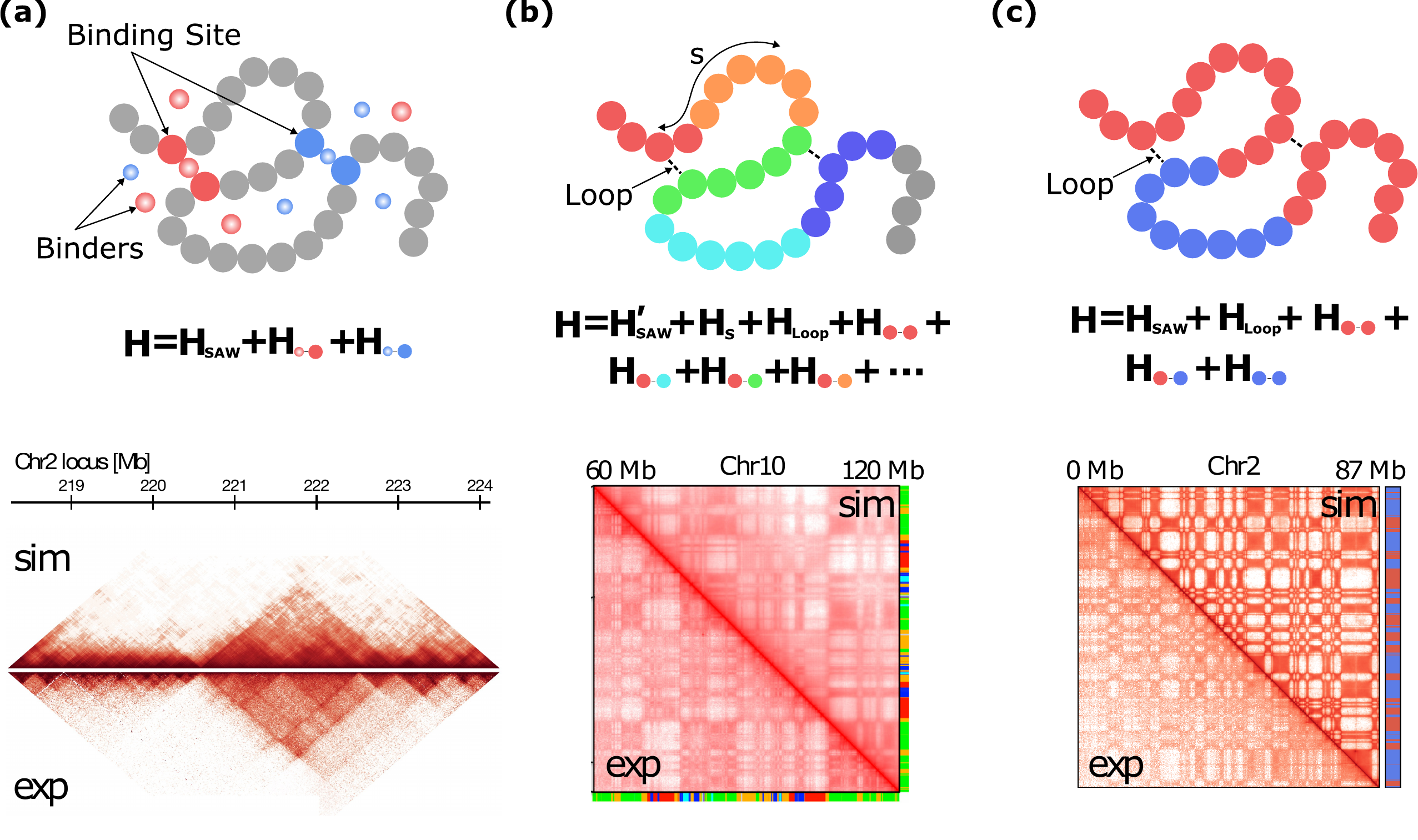}
\caption{Comparison of three polymer-based models for chromosomes: (a) SBS, (b) MiChroM, and (c) CCM. Schematics of the models (upper panel) along with the description of effective Hamiltonian, where different colors represent binders (binding sites) or locus types and $H_{\text{SAW}}$ ($H_{\text{SAW}}^{\prime}$) corresponds to the energy of a self-avoiding chain (with chain crossing).  The lower panels compare the predicted contact maps with Hi-C data. The contact maps in panels a and b are replotted using the data from Ref.~\citenum{bianco2018NatGene} and reproduced from Ref.~\citenum{DiPierro2016}, respectively.}
\label{fig:fig2}
\end{figure}

In the SBS model \cite{Barbieri12PNAS}, chromosome is modeled as a self-avoiding polymer in which a fraction ($f$) of sites could bind simultaneously to one or more diffusing molecules (such as CTCF proteins or transcription factors) with interaction strength, $E_X$ (Fig. \ref{fig:fig2}(a)). In the first application, $f$ was set to 0.5, and the binding sites were randomly selected. 
By varying the concentration, $C_m$, of binders it was found that there a transition to a globular state at a critical value of $C_m$.   By varying the three variables in the SBS model \cite{Barbieri12PNAS} $f$, $E_X$, $C_m$, and the fraction of compact conformations, they obtained the contact probability $P(s)$ as a function of genomic separation ($s$) which was in qualitative agreement with experiments \cite{lieberman09Science}. 
The SBS model was generalized \cite{bianco2018NatGene}  to describe the effects of structural variants (deletions, inversions, and duplications), which required varying  the  number of binding sites, $n$. In the  resulting  algorithm \cite{bianco2018NatGene}, the parameters in the model are iteratively adjusted till agreement with measured contact maps is obtained (Fig. \ref{fig:fig2}(a)). 

A model similar to SBS  \cite{brackley2016NAR}  described the role of transcription factors (TFs) on the folding of interphase chromosomes. The distinct epigenetic states (``active" and ``inactive") for the chromatin fiber were chosen to reflect the active and inactive states of the chromosomes. The TFs bind to the active and inactive beads on the chromosomes with different interaction strengths. 
By choosing different binding energies of the TFs to the active  and inactive loci, it was found that the calculated contact maps \cite{brackley2016NAR} were in fair agreement with experiments \cite{Rao2014Cell}. 


\subsection{Copolymer models}
The checkerboard-like pattern in the contact map of interphase chromosomes correlates  the genetic activity of given loci, which is interpreted to mean compartment formation on several Mb scale is a consequence of segregation between genetically active and inactive loci \cite{lieberman09Science,Rao2014Cell}. 
To explain the Hi-C data, copolymer models \cite{Jost14NAR,Zhang15PNAS,DiPierro2016,Michieltto2016PRX,Shi18NatComm} were proposed along with the effective  energy function of a polymer chain that includes attractive pairwise interactions between non-bonded loci that differ depending on the epigenetic states \cite{Jost14NAR} (Fig.~\ref{fig:fig2}). 

\subsubsection{MiChroM}
The  MiChroM (Minimal Chromatin Model) \cite{DiPierro2016} was introduced (Fig.~\ref{fig:fig2}(b)) to understand the details of the structural ensemble of interphase chromosomes. The   energy function incorporates  contact interactions for six  loci types \cite{Rao2014Cell}. Terms for CTCF loop interactions ($H_{\text{Loop}}$), and local compaction term ($H_s$) that depends on genomic distance from a given locus were also included, in addition to the self-avoiding interactions that allow for limited chain crossing ($H_{\text{SAW}}^{\prime}$). 
The energy function with twenty-seven parameters is optimized through the maximum entropy procedure \cite{Zhang15PNAS}. The simulations using the resulting transferable model  quantitatively reproduced  experimental contact maps for human chromosomes \cite{DiPierro2016}.


\subsubsection{Chromosome Copolymer Model (CCM)} 
The CCM is a copolymer model that faithfully captures organizational and dynamical features of chromosome \cite{Shi18NatComm}. The interphase chromosomes are represented as self-avoiding copolymers consisting of  A and B loci (Fig.~\ref{fig:fig2}(c)).  The A (B) type locus corresponds to the euchromatin or active (heterochromatin or repressive) epigenetic  state. Interactions between the loci are modeled using   locus-dependent interaction strength.  
The CCM uses harmonic potential to model the presence loop anchors that are typically CTCF stabilized by cohesin \cite{Rao2014Cell}.   
By using the Flory-Huggins theory \cite{Flory42JCP,Huggins42JACS} the phase separation was mimicked by choosing a smaller  value for the interaction between A and B type loci. Simulations using the CCM quantitatively reproduce  the contact maps at both 1.2 kb and 50 kb resolution (Fig. \ref{fig:fig2}(c) shows one example).

\subsection{Phase separation as the mechanism of compartment formation}
Emergence of A/B compartments in Hi-C contact maps implies that  active chromatin loci are segregated from  inactive  loci \cite{Pollard2017CellBiology}. 
Copolymer  models  \cite{Barbieri12PNAS,Jost14NAR,DiPierro2016,michieletto2016PNAS,Shi18NatComm} reproduce the compartment features, suggesting that physical process underlying the compartmentalization is microphase separation \cite{Hildebrand2020, Misteli20Cell, Feric2021}, where chromatin sites with similar histone markers tend to interact through direct interactions between nucleosomes or through histone-binding proteins \cite{Cui2000,Funke2016,Shimamoto2017}. 
Such an interpretation is supported by the intrinsic ability of chromatin to phase separate in \textit{vitro} \cite{Gibson2019cell}. 
If the Flory-Huggins $\chi$ parameter is large enough to cause microphase separation between different types of chromatin, the compartments can be recovered in the contact map \cite{Shi18NatComm,shin2023effective}. 

Experimental studies show that the protein HP1$\alpha$ forms liquid droplets both \textit{in vitro} and \textit{in vivo} \cite{Larson2017Nature, Strom2017Nature, Sanulli2019Nature}. Similarly, the Polycomb Repressive Complex 1 (PRC1), which alters histone markers,  form droplets \textit{in vitro} \cite{Plys2019}. These findings reinforce the notion that heterochromatin loci cluster together through phase separation. Although there is no direct evidence for a similar process in euchromatin, it is possible that euchromatin loci self-assemble through transcription condensate formed by RNA polymerases and coactivators \cite{Cho2018,Sabari2018,Boija2018Cell}.

\section{From Hi-C Data to three dimensional chromosome structures} 
A number of data-driven computational models~\cite{2012Kalhor90,di2016hi,DiPierro2017PNAS,2018Hua915,abbas2019integrating,2020Qi1905,Perez-Rathke2020,Kumari2020,shi2021PRX,2021DiStefano25,2022Boninsegna938,Contessoto2023,shin2023effective} have been proposed to calculate the ensemble of 3D chromosome structures that would be consistent with Hi-C data. 

\begin{figure}[h]
\centering\includegraphics[width=0.9\linewidth]{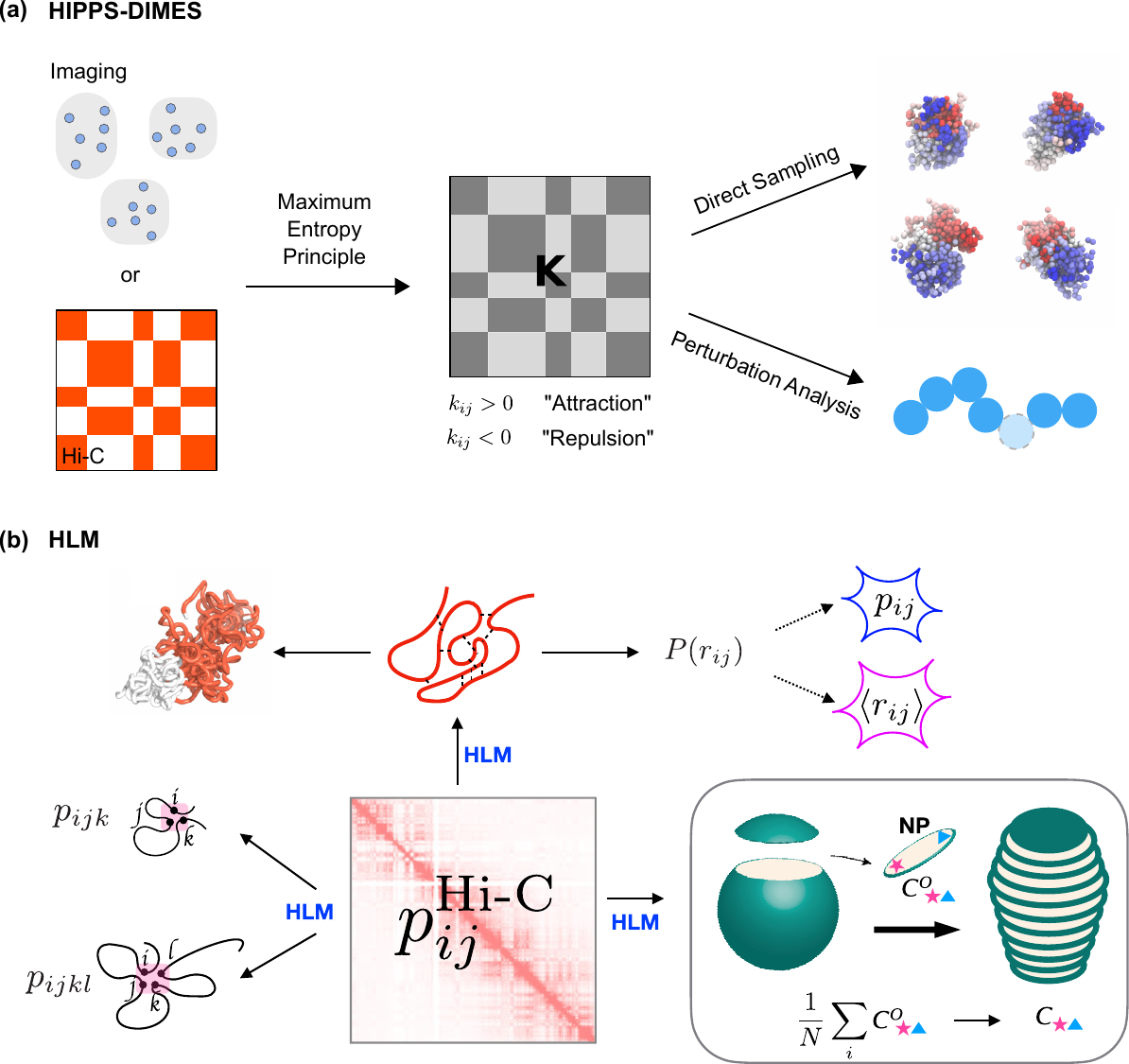}
\caption{(a) HIPPS-DIMES uses the experimental data (Hi-C contact map or distance map) as constraints to generate an ensemble of 3D structures. 
(b) Heterogeneous loop model (HLM) to extract structural features of 3D chromosome folding from Hi-C data. Three and four-way contacts can be enumerated.}
\label{fig:fig3}
\end{figure}

\subsection{HIPPS and DIMES}

HIPPS (Hi-C-polymer-physics-structures) \cite{shi2021PRX} and DIMES (Distance Matrix to Ensemble of Structures) \cite{Shi2023}, collectively referred to as HIPPS-DIMES from now on,  were created to utilize Hi-C contact maps or imaging data to generate three-dimensional coordinates of chromatin loci (Fig.~\ref{fig:fig3}(a)). The central idea, which  is similar to generative modeling in machine learning \cite{Jebara2004},  is to derive  a distribution function whose random variables are the three-dimensional coordinates of the chromatin loci.  HIPPS-DIMES utilizes the maximum entropy principle \cite{Jaynes1957,Press2013}, which produces a unique distribution that satisfies the constraints derived by the experimental data. 

\begin{equation}
    P^{\mathrm{MaxEnt}}(\{\boldsymbol{x}_i\})=\frac{1}{Z}\exp\left(-\sum_{i<j}k_{ij}||\boldsymbol{x}_i-\boldsymbol{x}_j||^2\right)
\end{equation}

\noindent where \(\boldsymbol{x}_i\) are the 3D coordinates of the chromatin loci, indexed by, \(Z\) is the normalization factor, and \(k_{ij}\) are the Lagrange multipliers, which are determined to ensure the distribution satisfies the constraints of mean squared pairwise distances. The elements, $k_{ij}$, form  the connectivity matrix $\boldsymbol{K}$ (Fig. \ref{fig:fig3}(a)), which means,

\begin{equation}
    a_{mn}=\int \mathrm{d}\{\boldsymbol{x}_i\}P^{\mathrm{MaxEnt}}(\{\boldsymbol{x}_i\})(\boldsymbol{x}_m-\boldsymbol{x}_n)^2
\end{equation}

\noindent where \(a_{mn}\) is the target mean squared pairwise distance between the \(m^{\text{th}}\) and \(n^{\text{th}}\) chromatin loci. If the  imaging data are used as constraints, \(a_{mn}\) can be directly calculated. For Hi-C contact maps, \(a_{mn}\) can be inferred  from the pairwise Hi-C contact probability, \(a_{mn} \sim c_{mn}^{-1/\alpha}\), where \(\alpha\) is a free parameter typically that ranges between 3 and 5, as suggested by polymer physics theory \cite{shi2019NCOMM,Shi18NatComm} and experimental measurements \cite{wang2016science}.


HIPPS-DIMES have been validated by quantitatively reproducing Hi-C contact maps and  mean distance maps across multiple  length scales from kbs to Mbs \cite{shi2021PRX,Shi2023,Jeong24eLife}, corresponding to the sub-TAD level to compartments across the entire chromosomes. Surprisingly,  DIMES \cite{Shi2023} reproduces the distribution of pairwise distances, even though only the mean is used as input, thus effectively capturing the variations in chromosome 3D structures. Since it makes no assumptions about the cell type, it has been applied to Hi-C contact maps from different stages of the cell cycle to investigate large-scale structural changes from interphase to mitotic chromosomes \cite{shi2021PRX, Dey23CellReports}.

 By interpreting the Lagrange multipliers \(k_{ij}\) as ``spring constants" and the maximum entropy distributions as Boltzmann distributions of a generalized spring network \cite{flory1976statistical} or generalized Rouse model \cite{doi1988theory, Bryngelson1996, Shukron2017pre}, the effect of perturbation of the  effective interactions on the 3D  chromosome structures may be assessed.  Unlike in the  generalized Rouse model \cite{shi2019NCOMM}, \(k_{ij}\) may be negative, representing ``repulsive" interactions between loci.  Analysis of single loci deletion  showed that the boundary loci that separate the A/B compartment are more important for retaining the structural integrity of chromosomes \cite{Shi2023}. 

\subsubsection{Single-cell chromosome structure and structural heterogeneity}

Using imaging techniques, several groups have quantified the degree of heterogeneity in chromosome structures.  By visualizing the centroids of multiple TADs spanning entire chromosomes in human cells \cite{wang2016science}, it was shown that compartments are preserved  even in individual cells. This finding confirmed that the A/B compartments pictured in Hi-C contact maps are indeed physically separated. Detailed analysis \cite{Shi18NatComm} of the imaging data \cite{wang2016science}  revealed that exact chromosome conformations exhibit a widespread continuous distribution.  A high-throughput experiment, which was used to obtain a large dataset of measurements of distance measurements for about hundreds of pairs of loci in human fibroblast cells \cite{finn2019Cell,Su2020}, showed that the distributions of distances between any pair of loci are broadly distributed. 
Together, these results show the presence of extensive heterogeneity in genome organization on both small (0.1 Mbps) and large (100 Mbps) length scales. Supplementary Figure 1(a) shows four examples of single-cell chromosome distance maps, which are visually distinct from each other. The HIPPS-DIMES model shows \cite{Shi2023} that the distribution of pairwise inter loci distances is well fit by a non-central chi distribution predicted from the generalized Rouse model (Supplementary Figure 1(b)) or, more specifically, the Redner-des Cloizeaux form for the distribution of distances between two monomers in a polymer chain \cite{shi2019NCOMM}. 
Single-cell measurements~\cite{stevens2017Nature,bintu2018Science} also showed that TADs are not conserved structural units but rather adopt different structures from cell to cell. Surprisingly, it was demonstrated  \cite{bintu2018Science}  that even upon  depletion of cohesins, TAD-like structures are observed in individual cells, although the preferential locations of TAD boundaries observed in ensemble Hi-C contact maps are not as transparent. Remarkably, HIPPS-DIMES method captures the variation of TADs structure in both WT and cohesin-depleted cells \cite{Jeong24eLife} using only the Hi-C data alone. 

\subsection{Heterogeneous loop model}
The heterogeneous loop model (HLM)~\cite{Liu2019BJ,Liu20NAR,liu2021extracting,liu2022dissecting} determines the structural origin of genome function by building 3D structural ensemble of chromatin from Hi-C data (Fig.~\ref{fig:fig3}(b)). 
The model is based on Gaussian random loop model~\cite{2007vanDriel051805}, introduced  when the theoretical research on the conformation of interphase chromosomes was burgeoning. 
The original random loop model~\cite{2007vanDriel051805} 
was based  on the ideal polymer chain with randomly placed loops along the locus of chromosomes. 
Similar approaches were made  to address the effects of close contacts on the size of a generic polymer chain~\cite{solf1995JPA,Bryngelson1996,zwanzig1997JCP}. 

HLM adapts the random loop model by placing pairwise intra-chromosomal contacts at specific locations along the genomic locus, so that they quantitatively reflect the contact information  in Hi-C map.  
Chromatin fibers are represented as a linear polymer chain composed of $N$ coarse-grained segments~\cite{2018Orland2286,2020Onami020,2020Onami2220,shi2021PRX}.  
The HLM~\cite{Liu2019BJ,Liu20NAR,liu2021extracting} uses a sum of harmonic restraints on the spatial distances between all the pairs of segments. 
Given the  potential of HLM, the probability of the chromatin to adopt a particular conformation, represented by the vector ${\bf r}$, is then written in a Boltzmann form. 
The strength of harmonic restraints 
is determined based on the Hi-C contact probability. 


Thanks to the simplicity of the formalism, HLM and its variants have been used to analyse experiments \cite{2018Orland2286, Liu2019BJ,Liu20NAR,liu2021extracting,2020Onami020}. 
An ensemble of chromosome conformations generated using HLM not only visualizes the chromosome structures but also allows  one to provide insights into the chromosome organization inferred by experiments measurements.  Furthermore, structural analyses have been efficacious in 
(i) the validation of FISH measurement on the spatial distribution of topologically associated domains (TADs) and visualization of phase-segregated A- and B-type TADs;  
(ii) understanding the structural origin of variable gene expression level in a particular gene domains of two distinct cell lines; 
(iii) description of structural changes along the cell cycle. Finally, despite the cell-to-cell variability of 3D genome over population \cite{shi2019NCOMM,shi2021PRX}, 
the intra-chromosomal inter-locus distance distributions predicted by HLM are in excellent agreement with measurements \cite{liu2021extracting,takei2021integrated}.


\subsubsection{Multi-way chromatin contacts from 3D structures} 
The importance of DNA looping is often discussed as the initiation step in gene expression. However, there is growing evidence that ``chromatin hubs” (assemblies of multiple genes and enhancers) are important. 
A number of experimental techniques, such as chromosome-walk~\cite{2016Tanay296}, 3way-4C~\cite{2016Tanay296}, Tri-C~\cite{Oudelaar2019Natcomm}, MC-4C~\cite{2018deLaat1151,2020deLaat364}, GAM~\cite{2017Pombo519}, and SPRITE~\cite{2018Guttman744},   identified multi-way chromosome interactions, deemed to be important in gene expression. 

Multi-way contacts can  be counted using  3D chromosome structures \cite{shi2021PRX,Shi2023,liu2021extracting}.  
The pairwise contact probabilities are not independent from each other, but are linked through certain types of correlations reflecting the underlying chromosome structure. The expression 
derived from HLM not only complements experimental identification of multi-way chromatin contacts, but can also be used to study cell-line dependent assemblies of chromatin hubs and condensates involving multiple CTCF anchors. 

\section{Statistical potentials for conventional and inverted nuclei}
 



The  statistical potential concept (SP) is well-known in  protein \cite{Tanaka76Macromolecules,Miyazawa96JMB} and RNA \cite{Dima05JMB} folding. 
The idea is that effective free energy of interaction between a pair of  amino acid residues may be determined using the frequency of native contacts between them in the folded structures. 
Recently, we showed that  SP is also useful in  extracting effective chromosome interactions from Hi-C data  \cite{shin2023effective}.

\subsection{Application to single chromosomes}
\begin{figure}[h]
\centering\includegraphics[width=\linewidth]{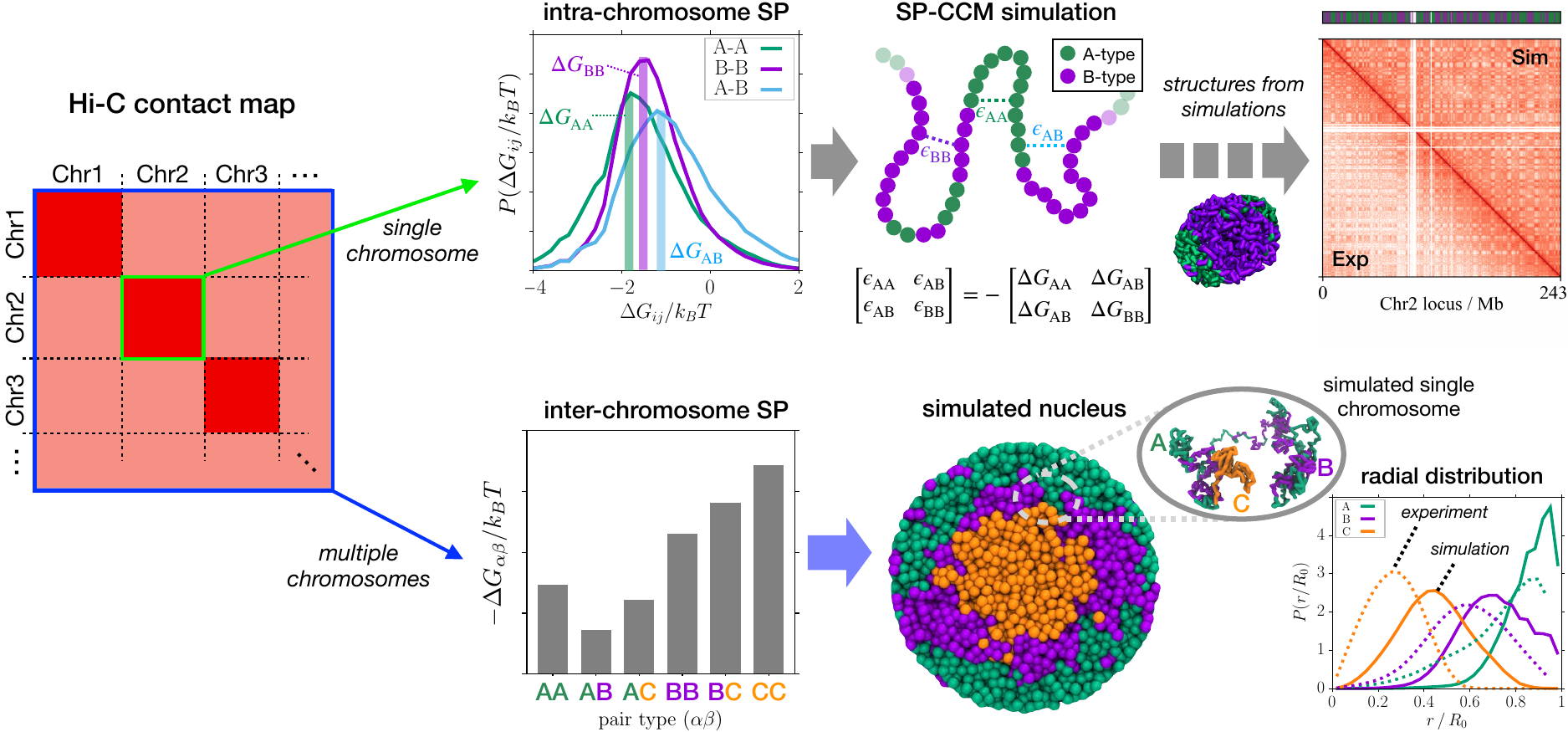}
\caption{Statistical potentials (SP) for chromosome interactions, extracted directly from Hi-C data, for use as effective interactions in polymer simulations. Based on the extracted SPs, the structural dynamics of a single chromosome (top) or multiple chromosomes (bottom) may be simulated. Graphics are reproduced from Ref.~\citenum{shin2023effective}.}
\label{fig:sp}
\end{figure}

The SP for a pair of specific loci, $i$ and $j$, on a single chromosome is defined as,
\begin{equation}
\Delta G_{ij} = - k_B T \ln \left[ \frac{P_\text{exp}(i,j)}{P_\text{ref}(i,j)} \right]~,
\label{eq:SP}
\end{equation}
where $P_\text{exp}(i,j)$ is the pairwise contact probability given by the Hi-C contact map. In Eq.~\ref{eq:SP}, $P_\text{ref}(i,j)$ is the contact probability expected in a reference system,  which is chosen to be an ideal homopolymer,  $P_\text{ref}(i,j) = |i-j|^{-3/2}$ \cite{doi1988theory}.
Thus, $\Delta G_{ij}$ gives the effective interaction for the contact pair, excluding the entropic contribution of the polymer backbone to the contact free energy. 
We demonstrated that when the SP is classified based on the locus type in a given pair (A-type for gene-rich loci; B-type for gene-poor loci), the mean values, $\Delta G_\text{AA}$, $\Delta G_\text{BB}$, and $\Delta G_\text{AB}$, could be used as effective interactions in polymer simulations \cite{shin2023effective}.  
Remarkably, the mean values yield the Flory-Huggins parameter, $\chi_\text{FH} = [\Delta G_\text{AB} - (\Delta G_\text{AA} + \Delta G_\text{BB})/2 ]/k_B T$, which is positive, thus ensuring that A and B loci undergo microphase separation. 
The CCM simulations, which take $\epsilon_{\alpha\beta} = - \Delta G_{\alpha\beta}$ ($\alpha, \beta = \text{A}$ or B) for the energy depth of the non-bonding potentials, generate a structural ensemble that is in excellent agreement with experimental results (Figure~\ref{fig:sp}, top). 
When the CTCF-mediated loop anchors are taken into account, the SP-based CCM (SP-CCM) simulations also capture the TADs, which appear as contact-enriched regions along the diagonal of a Hi-C contact map. 

\subsection{Application to multiple chromosomes}

If one is focused on the organization of a single chromosome, the mean SP values of individual chromosomes differ depending on the chromosome length, \emph{i.e.}, $\Delta G \sim - \ln L_\text{chr}$ since a longer chromosome requires larger free energy cost to be folded into compact structures \cite{shin2023effective}.
On the other hand, for a system of multiple chromosomes, such as the whole genome, the SP values for specific pair types are expected to be less distinguishable between chromosomes according to the quasi-chemical approximation \cite{Miyazawa85Macromolecules}.
Indeed, multi-chain copolymer models, whose non-bonding interaction parameters depend only on the epigenetic states of loci on individual chromosomes, have been successful in simulating the organization of  whole genomes \cite{Fujishiro2022,Contessoto2023}.

Notably, we demonstrated that the SP is also effective in capturing the interaction energies between specific locus pair types and simulating the spatial organization of chromosomes in inverted nuclei. 
Unlike typical interphase nuclei, where gene-rich euchromatin (A-type) and gene-poor heterochromatin (B-type) are located in the interior and near the periphery, respectively \cite{Misteli20Cell}, the nuclei of rod photoreceptor cells in nocturne mammals exhibit the opposite trend---euchromatin is located on the periphery whereas heterochromatin is found in the the interior \cite{Solovei2009}.  
In particular, the centromere and pericentromeric heterochromatin (collectively classified as C-type loci) are localized at the nuclear center. 
A previous simulation study \cite{Falk2019} showed that in order to obtain the desired pattern of nuclear compartments the energy scales should satisfy $\epsilon_\text{CC} > \epsilon_\text{BB} > \epsilon_\text{AA}$. 
The SPs extracted from interchromosome Hi-C data for inverted nuclei automatically satisfy the optimal condition so that the multi-chain  polymer simulations of mouse chromosomes using the SPs yield the spatial arrangement of A, B, and C type loci, which is in good agreement with the experimental results \cite{shin2023effective} (Figure~\ref{fig:sp}, bottom). 
On the other hand, it was  found \cite{Solovei2009} that additional interactions between heterochromatin and nuclear envelope (which is due to nuclear lamina) is needed to reproduce the organization of the conventional nuclei. 
Our SP method revealed that the effective interchromosome interactions implied in the Hi-C data for conventional nuclei do not differ significantly from those for inverted nuclei \cite{shin2023effective}. 
When we inferred the effective interactions between heterochromatin and nuclear envelope through the SP-like procedure from the experimentally measured radial distribution of chromosome loci and subsequently used those interaction energies for the input parameters of simulations, we obtained the spatial arrangement of chromosomes as observed in conventional nuclei. The successful applications of the SP concept, using only Hi-C data \textit{without} any free parameter, illustrate the power of the method.



\section{Mitotic Chromosome Structure: Random Helix Perversion}
\begin{figure}[H]
\centering\includegraphics[width=0.9\linewidth]{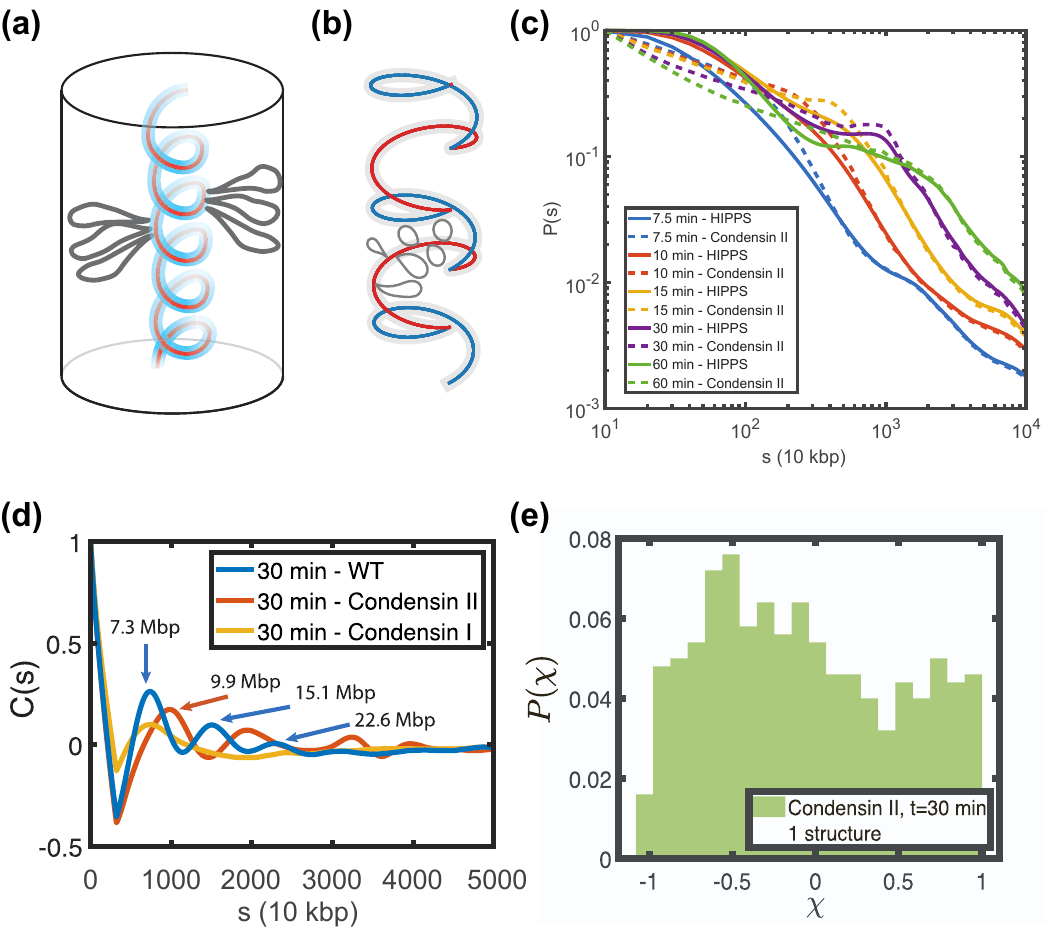}
\caption{(a) Cartoon of staircase model for mitotic chromosome \cite{Gibcus18Science}. (b) Schematic for helix perversion along the mitotic chromosome axes displaying half-helices with alternating  handedness \cite{Chu20MolCell}. (c) Comparison of the contact probability, $P(s)$, between experiments for DT40 condensing I-depleted cells and HIPPS prediction shows excellent agreement. (d) Angular correlation $C(s)$ (Eq. \ref{Cofs}) as a function of genomic distance. Comparison between WT, Condensin I-depleted (orange) and Condensin II-depleted (red) DT40 cells at 30 minutes. (e) Distribution of the local helical handedness parameter $\chi$ for HIPPS predicted structures in condensin II-depleted DT40 cells at $t=30$ mins. $\chi$ is defined as $\chi=(\vec{AB}\times \vec{CD})\cdot \vec{EF}/|\vec{AB}||\vec{CD}||\vec{EF}|$ where $\vec{AB}$ and $\vec{CD}$ are vectors between the $i^{th}$ and $(i+p/2)^{th}$ loci, and $(i+p/4)^{th}$ and $(i+3p/4)^{th}$ loci, respectively. The vector $\vec{EF}$ connects the midpoints of $\vec{AB}$ and $\vec{CD}$. Panels a and b are reproduced from Refs.~\citenum{Gibcus18Science} and \citenum{Chu20MolCell}, respectively. Panels c--e are reprinted from Ref.~\citenum{Dey23CellReports}.}
\label{fig:review_mitotic}
\end{figure}

It has been known for a long time that mitotic chromosomes, which are helical \cite{De88Cell,Earnshaw88Bioessays}, are highly condensed. Several models to explain the mitotic structures have been proposed \cite{Chu20MolCell,Chu20PNAS,Woodcock84JCB,Zhang16PRL,Gibcus18Science,Dey23CellReports} However, only recently the nature of the helix and the role the motor proteins, especially the roles of condensin I and condensin II, in shaping the shapes of mitotic chromosomes have been  elucidated \cite{Dey23CellReports,Sun18ChroRes}.  The theory based on the cooperative interaction between multiple condensins I and II motors showed that the latter extrudes loops that are roughly six times longer than the former \cite{Gibcus18Science,Dey23CellReports}. Although the ratio of the loop lengths predicted by the multiple motor model is similar to the equilibrium polymer simulations performed to fit the Hi-C data \cite{Gibcus18Science}, the absolute values of the loop sizes are different. In addition, the active motor model \cite{Dey23CellReports} predicts that the the loops, which are not arranged in a consecutive manner because of the presence of nested and the so-called Z-loops, are stapled to a dynamically changing central helical scaffold formed by the motors. This prediction differs from the interpretation using polymer simulations \cite{Gibcus18Science}, performed without motor activity, that the loops are consecutive, emanating from a central spiral staircase formed by the motors (Fig. \ref{fig:review_mitotic}(a)).

To shed light on the nature of the helical conformations in mitotic chromosomes, we used the Hi-C data at different stages of the cell cycle in DT40 chicken cells \cite{Gibcus18Science},  and calculated the time dependent changes in the 3D structures using the HIPPS method \cite{Dey23CellReports}. The calculated dependence of the contact probability as a function of the genome distance at various times in cells in the absence of condensin I  are in excellent agreement with experiments (Fig.~\ref{fig:review_mitotic}(c)). The peak in the $P(s)$, which is attributed to the emergence of helicity, moves to larger $s$ values at the cells approach mitosis. The spatial extent of helix formation is quantified using,
\begin{equation}
    C(s) = \langle \textbf{r}_{i,i+d}\cdot \textbf{r}_{i+s,i+s+d}\rangle
    \label{Cofs}
\end{equation}
\noindent where \textbf{$r_{i,j}$} is the vector between the $i^{th}$ and $j^{th}$ loci. For a perfect helix, $C(s)$ should be periodic with a period that is proportional to the pitch of the helix. The HIPPS calculations show that there is modest helix formation even in the absence of condensin II. Comparison of $C(s)$ at 30 min for the  WT, condensin I, and condensin II cells shows that the periodicity is more prominent when condensin I is depleted, thus supporting the finding that condensin II confers helicity to the dynamic scaffold \cite{Gibcus18Science} (Fig.~\ref{fig:review_mitotic}(d)). Using the HIPPS-generated 3D structures, we calculated the helix order parameter,  expressed in terms  of the angle $\chi$, that we use to extract the local handedness. The distribution, $P(\chi)$ for a single conformation at 30 min is broad with $\chi$ sampling both positive (right handed) and negative (left handed) helical conformations (Fig.~\ref{fig:review_mitotic}(e)). In other words the handedness changes along the mitotic axis, as first articulated by Klecker \cite{Chu20MolCell}, and is referred to as ``perversion",  across the chromosome, possibly even randomly \cite{Dey23CellReports}. Thus, mitotic chromosomes have all the signs of helix ``perversion" (Fig.~\ref{fig:review_mitotic}(b)) whose importance in mitosis has been elucidated in a series of insightful studies \cite{Chu20PNAS,Chu20MolCell}. 

\section{Dynamics of Interphase Chromosomes}

Because chromatin dynamics play a crucial role in all biological functions including gene expression, DNA repair, and many other essential functions, understanding the movement of chromosome loci within the confined space of the cell nucleus is of paramount importance. Techniques such as FISH and its derivatives (e.g., multiplex FISH), as well as the Hi-C method, cannot probe genome dynamics since they are conducted on fixed cells. Live-cell imaging techniques are necessary to investigate the dynamic aspects of genomes. At present, these techniques primarily rely on labeling chromatin loci through DNA-binding proteins \cite{Robinett1996} or recently through dCas9 \cite{Chen2013} with fluorescent tags, possibly at multiple sites. In the gene expression context, it is important to  simultaneously monitor promoter-enhancer dynamics and the gene product, as has been done recently \cite{Bruckner2023}. Althougth there are limitations, polymer physics provides a framework to interpret such experiments.

\subsection{Polymer Physics of Chromatin Dynamics}

Because above a certain length  chromosomes can be modeled as a polymer elucidating their dynamics  from a polymer physics perspective is natural, as are methods for determining their structures. 
The simplest  model is an ideal or Rouse chain  in which all interactions except chain connectivity are neglected. The Rouse model \cite{doi1988theory} shows that the mean square displacement (MSD) of a monomer behaves as $\mathrm{MSD} \sim D t^{1/2}$ for small $t$, and exhibits normal diffusion ($\mathrm{MSD} \sim t$) at long times. Generally, the monomer MSD  can be written as $\mathrm{MSD} \sim D t^{\alpha}$, where $\alpha$ is the diffusion exponent and $D$ is the diffusion coefficient. If $\alpha < 1$, the process is  sub-diffusive, and if $\alpha > 1$, it is super-diffusive. Both cases are sometimes referred to as anomalous diffusion.

It is expected  that the structure of chromatin should determine its global dynamics \cite{Liu2018PLoSCB}, as is the case for polymers \cite{deGennesbook}. This connection can be illustrated using the following scaling argument. The characteristic relaxation time of a polymer segment, $\tau_r$, is given by $\tau_r = R^2 / D$, where $R$ is the  length scale of the segment, and $D$ is the diffusion coefficient of the center of mass (COM) of the segment. The dynamics of monomers are sub-diffusive for $t \ll \tau_r$ and exhibit normal diffusion for $t \gg \tau_r$. For an ideal chain in which correlation effects are neglected, all monomers contribute to the diffusion of the COM. Thus, $D$ must scale as $N^{-1}$, where $N$ is the number of monomers. In contrast, consider a compact globular structure formed by a single polymer. Assuming that its internal motion is sluggish, so the chain would move as a rigid body on the time scale of $\tau_r$. Under these conditions, $D \sim N^{-2/3}$ because only the surface monomers contribute to the diffusion of the whole chain. Generally, we expect $D \sim N^{-\theta}$, where $\theta$  quantifies the number of monomers that contribute to the global motion of the COM. 
Using the relation $R \sim N^{2\nu}$, where $\nu$ is the Flory exponent, we obtain $\tau_r \sim N^{2\nu + \theta}$. The length scale associated with monomer diffusion at time $\tau_r$ must coincide with the length scale of the chain itself, leading to $\tau_r^{\alpha} \sim R^2 \sim N^{2\nu}$. Hence, the diffusion exponent of a monomer for time $t < \tau_r$ is given by $\alpha = 2\nu / (2\nu + \theta)$. For an ideal chain,  $\nu = 1/2$ and $\theta = 1$, and $\alpha = 1/2$, which aligns with the predictions from the Rouse model. For the  fractal globule \cite{grosberg1993EPL}, which has been proposed to be the model for chromatin structure \cite{Mirny2011},  $\alpha = 0.4$ (with $\nu = 1/3$ and $\theta = 1$). 

The distribution of the distance between two monomers along a polymer chain  is given by $P(r) \sim r^{2+g} \exp(-Br^{\delta})$, where $B$ is a constant depending on the specific polymer model \cite{Redner1980, Jannink1990}, $g$ is the ``correlation hole” exponent, and $\delta$ is related to the Flory exponent $\nu$ by $\delta = 1/(1 - \nu)$. The exponents $g$ and $\delta$ govern the small and long length scales of polymer conformation, respectively. Since the diffusion exponent of monomers on the small time scale is directly related to its structural characteristics, the Flory exponent $\nu$ affects the dynamics at all times. Moreover, the dynamics of contact formation between two monomers are connected to both $g$ and $\nu$. Based on the arguments given previously  \cite{Toan2008}, combined with the scaling argument $\alpha = 2\nu / (2\nu + \theta)$, it can be shown that the compactness of the exploration of the space between two monomers before they come into contact can be quantified by the parameter $\gamma = (3+g)\nu / (2\nu + \theta)$, in three   dimensions. When $\gamma > 1$, the exploration is non-compact, and the mean first passage time of contact formation, $\tau_c$, is $\tau_c \sim N^{\nu(3+g)}$. For $\gamma < 1$, the compact exploration of the conformations between two monomers leads to $\tau_c \sim N^{2\nu + \theta}$. The dynamics of contact formation \cite{Amitai2018pre} are particularly important because they  control  enhancer-promoter (E-P) interactions, where the enhancer and promoter form direct contacts or come into proximity \cite{Schoenfelder2019} to induce the gene expression. Hence, the physics of search process between two distal genomic loci has direct biological consequences \cite{Bruckner2023}. Interestingly, the arguments given in the section, based on polymer physics, is most useful in elucidating some aspects of chromosome dynamics, which we discuss in the following sections.

\begin{figure}[H]
\centering\includegraphics[width=0.9\linewidth]{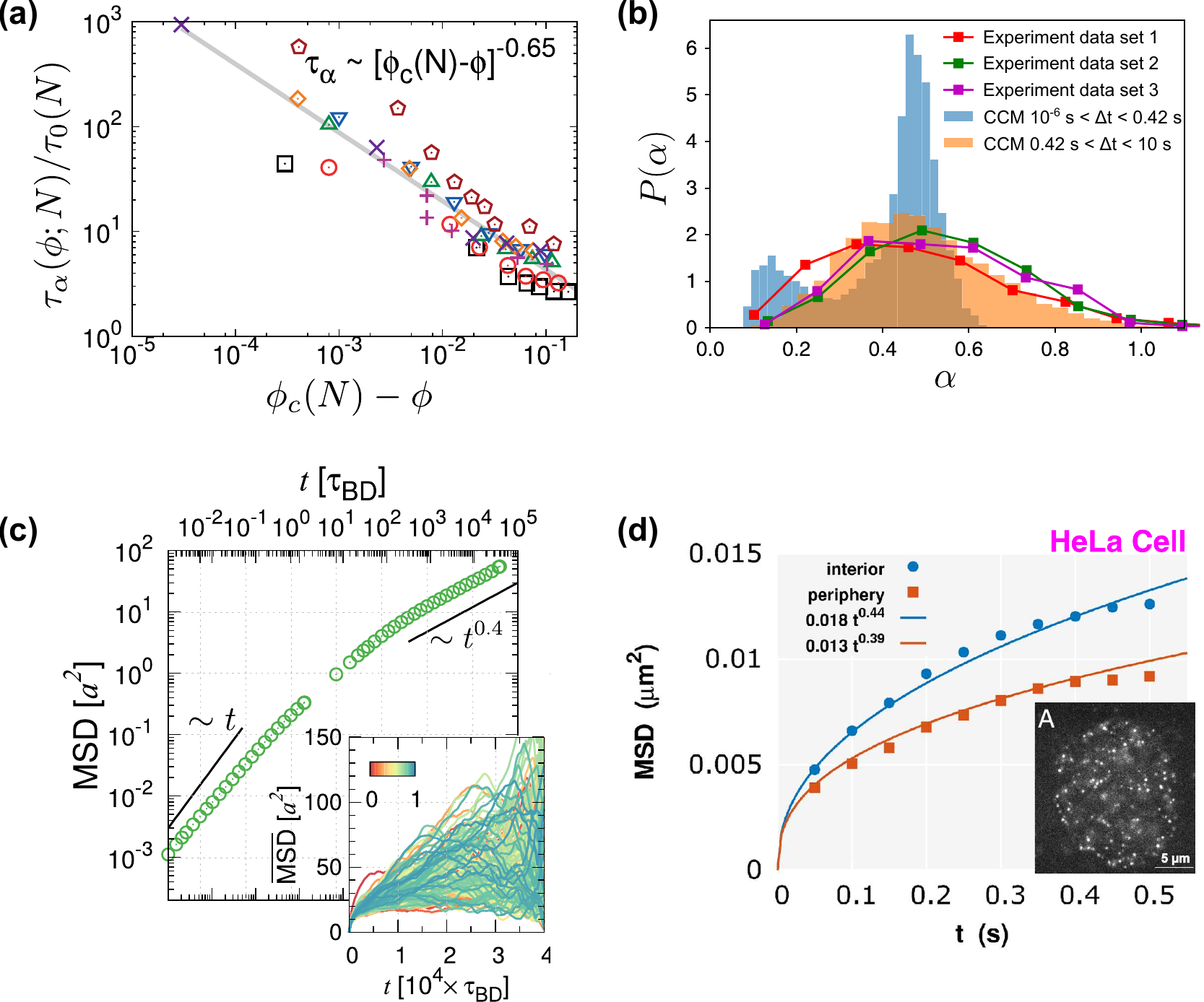}
\caption{(a) Confinement induced glassy dynamics. The power law divergence of relaxation time as the volume fraction approaches the critical value $\phi_c\approx 0.44$ \cite{2015Hyeon198102}. (b) Interloci interaction induced glassy dynamics in the CCM. Single-loci diffusion exponent $\alpha$ exhibits bimodal distribution at short times, and widespread distribution at longer times, which \textit{quantitatively} agrees with the experiment measurements \cite{bronshtein2015NatComm}. (c) Polymer model with fractal globule-like organization has a diffusion exponent of 0.4. (d) Mean Square Displacements (MSDs) measured for loci dynamics of HeLa cell lines \cite{shinkai2016dynamic}. Chromatin loci at the periphery of cell nucleus has diffusion exponent of $\sim 0.4$. The plots of panels a--d are reprinted or adapted from Refs.~\citenum{2015Hyeon198102}, \citenum{Shi18NatComm}, \citenum{Liu2018PLoSCB}, and \citenum{shinkai2016dynamic}, respectively.}
\label{fig:review_dynamics_fig}
\end{figure}

\subsection{Experimental measurements on chromatin loci dynamics}
Early experimental investigations into interphase chromosome dynamics revealed that individual loci exhibit constrained sub-diffusive motion in living cells on a timescale of up to a hundred seconds \cite{Marshall1997}. The length scale of the confinement is on the order of sub-microns, significantly smaller than the size of the nucleus. The authors suggest that this constrained motion results from the tethering of chromatin loci. Hence, the confinement size only depends on the distance between two tethering points, which can be much smaller than the size of the chromosome territories and the nucleus. Note that the sub-diffusivity of chromatin loci is not surprising since it may be explained by the polymeric nature of chromosomes. The exact value of the diffusion exponent on the other hand is informative as it can be used to differentiate between different models of chromatin dynamics. Recent experiments have shown that the diffusion exponent, $\alpha$, for chromatin is between 0.4 and 0.6 \cite{Levi2005, bronstein2009PRL, weber2010PRL, Chen2013, hajjoul2013GenomeRes, shinkai2016dynamic, Amitai2017cellreport, Gabriele2022, Mach2022, Bruckner2023}. A value of 0.4 is predicted by the fractal globule model, while 0.5 is predicted by the Rouse model. The difference between values 0.4 and 0.5 is small, and the statistical resolution of experimental measurements is often insufficient to confidently determine $\alpha$. It is likely that genome organization does not conform strictly to any generic polymer model but  adopts complex local structures with heterogeneous dynamic properties. Nevertheless, the finding $\alpha$ is found to be around 0.5 in most experiments indicates that chromatin loci dynamics is a consequence of generic polymer effects.

A biologically significant question arises in the context of the dynamics of promoters and enhancers and their communication during transcription regulation. Two recent experiments \cite{Chen2018,Bruckner2023}  studied promoter-enhancer dynamics using single-molecule tracking. The use of   multi-color labeling to monitor the enhancer, promoter, and transcription activity simultaneously in live cells \cite{Chen2018} provided evidence that promoter-enhancer physical proximity ($<200\ \mathrm{nm}$) is coupled with the transcription of the targeted gene in real time. More importantly, it was shown  that the contact is not stable but transient, it  breaks and reforms dynamically. Building upon this study,  the motions of pairs of DNA loci (promoter-enhancer pairs) of varying genomic separation were directly probed \cite{Bruckner2023}. It was found that the relaxation time between the pairs of loci depends on their genomic distance, and changes  with a power law exponent close to 0.7. This value is surprising, as it is much lower than what would be predicted for a compact structure using scaling arguments. The Rouse model predicts an exponent of 2, and the crumpled globule model predicts a value of 3/2. The smaller value found in experiments \cite{Bruckner2023}  implies that it takes much less time for two distal chromatin loci to come into physical proximity, possibly having crucial implications for gene regulation. A complete theoretical explanation for this important finding is lacking.

\subsection{Non-equilibrium glassy dynamics in interphase chromosomes}
In dramatic contrast to lower organisms, such as bacteria and yeast, chromosomes of mammals and humans are organized into self-similar fractal globule-like non-equilibrium structures.
Studies based on the reptation argument~\cite{grosberg1988JP,Rosa2008PLoSCB}, $\tau_{\rm rep}\sim N^{3.4}$~\cite{deGennesbook}, predicted that the time scale of equilibration of the entire human genome ($N\sim 6\times 10^9$ bp), which is greater than yeast ($N\sim 10^8$ bp) by 60 fold, is prohibitively long ($\tau_{\rm rep}^{\rm human}\sim 10^6\times \tau_{\rm rep}^{\rm yeast}$). According to an earlier estimate, the time of scale needed for human genome to come to equilibrium is around 500 years!~\cite{Rosa2008PLoSCB}.
As long as equilibration time is far greater than the cell doubling time,  chromosomes would be out of equilibrium in the cell nuclei. 
 
Although the arguments given above are plausible,  the origin of the vastly different time scales in the dynamics of chromosome organization between species is  unknown.  In particular, it is unclear if the differences arise due the action of specific  protein mediated genomic interactions or  confinement effects due to cell size variations alone is sufficient to explain the produce fractal globule structures and the dynamics. 
To address these issues, we~\cite{2015Hyeon198102} adopted Occam’s razor philosophy by modeling chromosome inside a nucleus as a self-avoiding chain, and performed Brownian dynamics simulations in a spherical cavity of varying cell sizes. 
Theoretical analysis based on simulations showed that the dynamical arrest of chromosomes in  tight confinement could be the origin of the major the difference in chromosome organizations between various species and cell type. 
Intermediate scattering functions of polymer chain established that the relaxation dynamics of the chain decreases dramatically  as the polymer volume fraction inside the sphere  approaches a critical value associated with the onset point of dynamical arrest. In particular, we showed that $\tau(\phi;N)\sim (\phi_c(N)-\phi)^{-\nu_c}$ (Fig.~\ref{fig:review_dynamics_fig}(a)).
Using finite size scaling analysis, the critical volume fraction for dynamical arrest was determined to be $\phi_c\approx 0.44$ for a long chain limit ($N\gg1$). 
This study \cite{2015Hyeon198102} highlighted the importance of the dynamical origin of  territorial organization in human chromosomes inside nucleus and why each chromosome maintains fractal-like domains. In contrast, chromosomes of budding yeast are predicted to exhibit fluid-like properties ~\cite{2015Hyeon198102}.

In addition to the purely entropic confinement effect on genome organization, nucleosomes are known to interact with each other, which introduces an enthalpic effect \cite{Liu2024}. Assuming that chromatin loci associated with different epigenetic markers have preferential interactions, the CCM model was employed to investigate how inter-loci interaction strength influences the spatiotemporal properties of human interphase chromosomes \cite{Shi18NatComm}. When the interaction strength is comparable to thermal energy , chromosomes exhibit fluid-like behavior. Conversely, with strong inter-loci interactions ($= 2.4 k_BT$), glassy dynamics are predicted, characterized by the stretched exponential decay of the dynamic scattering function. Interestingly, recent single-molecule experiment \cite{Cui2000, Funke2016} and simulation studies \cite{Moller2019} suggests that average nucleosome-nucleosome interaction strength is between 2.0 to 3.0 $k_BT$. Notably, the glassy dynamics also predict significant variations in single-loci diffusion coefficients and exponents. At short timescales ($<1 \mathrm{s}$), chromatin loci display a bimodal distribution of diffusion exponents, indicating the presence of fast and slow populations. At longer timescales ($>1 \mathrm{s}$), the distribution of diffusion exponents becomes monomodal, aligning quantitatively with experimental measurements (Fig.~\ref{fig:review_dynamics_fig}(b)).


\subsection{Chromosome dynamics using heteropolymer models}
To go beyond the predictions based on polymer models  discussed above, it is profitable to use heteropolymer chromatin models  which incorporate  epigenetic (sequence)  information. Such models allow one to explore the spatiotemporal dynamics of chromosomes in a more realistic fashion~\cite{Shi18NatComm, Liu2018PLoSCB}. 

Interphase chromosome structures generated from the heteropolymer models fold hierarchically, described by the formation of TAD-sized chromatin droplets  \cite{Shi18NatComm}, which merge as compaction proceeds by the  Lifschitz-Slyozov mechanism. In the folding process, checkerboard pattern found in the Hi-C map is reproduced accurately \cite{Shi18NatComm,Liu2018PLoSCB}, which implies that the space-filling organization is the origin of spatio-temporal hierarchy manifested in the loci dynamics of chromatin. 
Brownian dynamics simulations of chromosome model was used  to map the time scale of chromatin dynamics onto the physical times~\cite{Liu2018PLoSCB}. 
Reflecting the hierarchical organization of chromatin fiber in chromosome, the local chromosome structures, exemplified by TAPs ($\sim$ 0.1--1 Mb), display fast dynamics with relaxation time of $<$ 50 sec, whereas the long-range spatial reorganization of the entire chromatin ($\gtrsim \mathcal{O}(10^2)$ Mb) occurs in hours, which is comparable to cell doubling time. The simulations also provided  the dynamic basis of cell-to-cell variability, and large length scale coherent motion~\cite{Liu2018PLoSCB,Shi18NatComm} found in experiments \cite{Zidovska2013}. 

The role of active forces in live cells, which are characterized by their vectorial nature at small scales~\cite{bruinsma2014BJ}, requires scrutiny.  
To understand their effects on chromosome dynamics, 
an active force maybe be modeled using stronger isotropic white noise~\cite{smrek2017small}, 
such that they act on the active loci along the chromatin fiber, whose genomic location is taken from epigenetic markers. 
We  showed \cite{Liu2018PLoSCB}  that although the overall motion of chromatin loci is accelerated, the subdiffusive behavior characterized by the dynamic scaling exponent of MSD $\sim t^{0.4}$ is  unchanged (Fig.~\ref{fig:review_dynamics_fig}(c)). Using the intermediate scattering function, the active forces were shown to accelerate the relaxation dynamics of chromatin domain described by the low frequency modes and at long time scales, but do not significantly change the local dynamics, corresponding to the high frequency modes, at short time scale. The finding that  the predictions using polymer physics dynamical scaling, MSD$\sim t^{2\nu/(2\nu+1)}\sim t^{0.4}$ continue to hold~\cite{weber2010PRL,tamm2015anomalous,shinkai2016dynamic,Liu2018PLoSCB} (see Fig. \ref{fig:review_dynamics_fig}(d) for one example), shows that the space-filling and hierarchical chain organization, associated with the exponent $\nu=1/3$ for fractal globules, is  the governing factor in the dynamics even in the presence of active force. Notably, this result is in stark contrast to the analysis of yeast cell that demonstrated MSD$\sim t^{0.5}$, associated with the exponent $\nu=1/2$ for equilibrium globules~\cite{hajjoul2013GenomeRes}.  

Simulations using the MiChroM model \cite{DiPierro2018a} found  a smaller diffusion exponent, $\alpha \approx 0.3$, for human interphase chromosomes, which has been reported for the osteosarcoma cell line U2OS \cite{bronstein2009PRL}. Analysis of Rouse modes in MiChroM reveals deviations from the traditional Rouse chain model, indicating the coupling of relaxation across different time scales. The velocity correlation functions \cite{Lampo2016BJ} reveal dynamically associated domains in which chromatin loci move coherently as found in other studies \cite{Liu2018PLoSCB,Shi18NatComm,Zidovska2013}.



\subsection{Transcription and Chromatin Dynamics}

Nowhere is dynamics more relevant than in transcription.  In eukaryotes, it is a well orchestrated process involving energy consumption  as well as  interactions with various proteins bound onto DNA. 
In the last decade,   several live-cell imaging experiments  reported on the effect of transcriptional activity on chromatin dynamics. 
It was observed that chromatin motion in human cells is suppressed during active transcription and enhanced upon inhibiting transcription \cite{Zidovska2013,Nagashima2019,Shaban2020} (Supplementary Figure 2 (a)-(b)). 
This observation is counter-intuitive because actively transcribing RNA Polymerase (RNAP) molecules are known to exert forces on DNA which should make chromatin more dynamic \cite{Yin1995, Wang1998}. 
In contrast, other experiments reported that gene-regulatory sequences of DNA are accelerated upon transcription activation whereas they become less dynamic upon transcription inhibition \cite{Gu2018} (Supplementary Figure 2 (c)).


In order to explain these contradictory findings, we performed simulations of CCM by including active forces which mimic the effect of active RNAP on chromatin \cite{Shin2024}.
The CCM chain, in the absence of active forces, exhibits A/B compartmentalization, and glass-like dynamics that are the intrinsic organizational and dynamical features of interphase chromosomes \cite{Shi18NatComm}. 
We found that the simulated MSD of chromatin decreases at a specific range of activity which is consistent with the force magnitude measured from single-molecule experiments \cite{Yin1995, Wang1998} (Supplementary Figure 2 (d)).
The decrease in MSD is also in good agreement with the experimental results, which is surprising since our model does not have  adjustable parameters. 
Such a decrease in the dynamics is induced by a transient disorder-to-order transition driven by  active forces \cite{Shin2024}. 
We demonstrated the plausibility of this mechanism by calculating the effective potential between the active loci whose minimum location becomes commensurate with the distance preferred by the attractive non-bonding potential (Supplementary Figure 2 (e)). 
The results hold as long as the density of active force along the polymer chain is  high. When we reduce the force density, there is a  regime (Supplementary Figure 2 (d)) in which the simulated MSD increases in the presence of active forces, which has also been observed  \cite{Gu2018}.


\section{Conclusion \& Outlook} 
In this short perspective, we have focused only on the organization and dynamics of chromosomes, which is a speckle in the vast subject of genome biology. Despite this caveat,  it is worth remembering that there has been extraordinary progress in our understanding of genome folding in the last two decades. We outline a biased view of some of the issues that could be addressed in  the coming years.  
(i) To the extent genome structure controls gene expression, the wealth of data that is being generated by Hi-C experiments in normal and cancer cells provides unique opportunities to correlate structure and function using some of the methods outlined here. In particular, it should be possible to use Hi-C data to explain the structural origin of enhanced expression of certain genes (see for example \cite{Rhie19NatComm}). 
(2)It is now clear that compaction of interphase chromosomes and the formation of mitotic structures with helical perversion require motor proteins (cohesin and condensin) \cite{Hirano97Cell,Uhlmann16NatRevMolCellBiol}. Although theories of how these motors extrude long loops, which is thought to be a first step in creating the distinct structures during the cell cycle, have been advanced \cite{Alipour12NAR,Fudenberg16CellReports,Takaki21NatComm}, they are not complete. Unlike our understanding of motors that transport cargo in cells \cite{Mugnai2020RMP}, relatively little is known about how the catalytic cycle of these motors,  and more importantly how multiple motors \cite{Dey23CellReports} affect chromatin dynamics  to fold genomes. 
(3) A major challenge is to probe, both in space and time, the key events that arise during transcription. To achieve this goal, one has to, at the minimum,  simultaneously monitor the interactions between transcription factors, polymerase, and regions in the chromosome carrying enhancer and promoters. This would require developments of new experimental techniques and theoretical concepts. Progress is being made \cite{Bruckner2023}, but one has a long way to go before quantitative understanding emerges.\newline

\textbf{SUMMARY POINTS}
\begin{enumerate}
\item A number of computational methods using polymer presentation of chromosomes have been successful in generating three-dimensional structures that reproduce the salient experimental features. 
\item Novel approaches that utilize Hi-C or imaging data have been invented to calculate three-dimensional structures interphase  chromosomes. The structures reveal the importance of multiway contacts. 
\item Mitotic structures, determined using the HIPPS method, exhibit helical perversion in which handedness changes along the mitotic axis. 
\item Effective interactions between chromatin loci can be extracted using the statistical potential concept introduced in protein and RNA folding. The use of such interactions in polymer simulations describes the structures of normal and inverted nuclei.
\item Simple polymer models are useful in teasing out generic aspects of chromosome dynamics. The models suggest that yeast chromosomes are ergodic where as mammalian chromosomes exhibit out of equilibrium glass-like dynamics.
\item The global structure, which depends on the genomic scale, determines the mean square displacement of the loci. Strikingly, the single loci diffusion exponent exhibits a broad distribution, ranging from small values to unity.
\end{enumerate}


\section*{DISCLOSURE STATEMENT}
 The authors are not aware of any affiliations, memberships, funding, or financial holdings that
might be perceived as affecting the objectivity of this review. 

\section*{ACKNOWLEDGMENTS}
We thank Mario Nicodemi, Andrea M. Chiariello, and Mattia Conte for providing useful discussion and their simulation data. This work was supported by a grant from the National Science Foundation (CHE 2320256) and the Welch Foundation through the Collie-Welch Chair (F-0019).

\newpage
%
\bibliographystyle{ar-style3.bst}

\end{document}